\documentclass[sn-nature,pdflatex]{sn-jnl}

\usepackage{graphicx}
\usepackage{multirow}
\usepackage{amsmath,amssymb,amsfonts}
\usepackage{amsthm}
\usepackage{mathrsfs}
\usepackage[title]{appendix}
\usepackage{xcolor}
\usepackage{textcomp}
\usepackage{manyfoot}
\usepackage{booktabs}
\usepackage{algorithm}
\usepackage{algorithmicx}
\usepackage{algpseudocode}
\usepackage{listings}

\begin{document}

\title[Automated Extraction of Heritage Material Constitutive Models]{Automated Extraction of Mechanical Constitutive Models from Scientific Literature using Large Language Models: Applications in Cultural Heritage Conservation}

\author[1]{\fnm{Rui} \sur{Hu}}
\equalcont{These authors contributed equally to this work.}

\author[1]{\fnm{Yue} \sur{Wu}}
\equalcont{These authors contributed equally to this work.}

\author[1]{\fnm{Tianhao} \sur{Su}}

\author[1]{\fnm{Yin} \sur{Wang}}

\author*[1]{\fnm{Shunbo} \sur{Hu}}
\email{shunbohu@shu.edu.cn}

\author[1]{\fnm{Jizhong} \sur{Huang}}

\affil[1]{\orgname{Shanghai University}, \orgaddress{\city{Shanghai}, \country{China}}}

\abstract{
The preservation of cultural heritage is increasingly transitioning towards data-driven predictive maintenance and "Digital Twin" construction. However, the mechanical constitutive models required for high-fidelity simulations remain fragmented across decades of unstructured scientific literature, creating a "Data Silo" that hinders conservation engineering. To address this, we present an automated, two-stage agentic framework leveraging Large Language Models (LLMs) to extract mechanical constitutive equations, calibrated parameters, and metadata from PDF documents. The workflow employs a resource-efficient "Gatekeeper" agent for relevance filtering and a high-capability "Analyst" agent for fine-grained extraction, featuring a novel Context-Aware Symbolic Grounding mechanism to resolve mathematical ambiguities. Applied to a corpus of over 2,000 research papers, the system successfully isolated 113 core documents and constructed a structured database containing 185 constitutive model instances and over 450 calibrated parameters. The extraction precision reached 80.4\%, establishing a highly efficient "Human-in-the-loop" workflow that reduces manual data curation time by approximately 90\%. We demonstrate the system's utility through a web-based Knowledge Retrieval Platform, which enables rapid parameter discovery for computational modeling. This work transforms scattered literature into a queryable digital asset, laying the data foundation for the "Digital Material Twin" of built heritage.
}


\maketitle

\section{Introduction}\label{sec:intro}

The preservation of cultural heritage is increasingly embracing digital technologies to enhance both understanding and protection of irreplaceable historical assets~\cite{jouan2020digital}. The concept of ``Digital Twin''---a dynamic virtual replica that mirrors the physical state and behavior of a real-world entity---has emerged as a transformative paradigm in heritage conservation~\cite{angjeliu2020development, pepe2021scan}. Coupled with ``Preventive Conservation,'' which emphasizes proactive monitoring and intervention rather than reactive restoration, these approaches promise to revolutionize how we safeguard our built heritage for future generations~\cite{dellatorre2021preventive}.

However, a Digital Twin is more than just a high-fidelity 3D geometric scan; it requires faithful physical simulation to predict structural responses under stress. Central to this physical fidelity is numerical simulation, particularly Finite Element Analysis (FEA), which enables researchers to assess structural safety and predict long-term behavior~\cite{roca2010structural, d2020modeling}. From the seismic vulnerability assessment of medieval cathedrals to the thermal stress analysis of ancient frescoes, numerical modeling has become an indispensable tool~\cite{betti2011numerical, pineda2017environmental}.

Yet, the reliability of any such simulation---and by extension, the predictive utility of the Digital Twin---is fundamentally constrained by the accuracy of its input parameters. Without accurate mechanical constitutive models to describe how materials respond to applied loads, even the most sophisticated Digital Twin remains a visual shell, incapable of reliable structural prognosis~\cite{vuoto2024shaping, funari2021parametric}.

Heritage materials present unique and formidable challenges for numerical modeling that distinguish them from modern engineering materials. Ancient mortars, weathered stones, decayed timber, and historical bricks exhibit significant variability arising from multiple sources: heterogeneous raw materials available to historical craftsmen, diverse manufacturing techniques that varied across regions and eras, centuries of environmental exposure and degradation, and the inherent complexity of aging processes~\cite{valek2013characterisation, siegesmund2014stone}.

Over the past several decades, substantial experimental research has been devoted to characterizing these materials. Studies have investigated the mechanical properties of Roman concrete~\cite{jackson2014mechanical}, the stress-strain behavior of Byzantine masonry~\cite{binda2012characterization}, the creep and relaxation of historical timber structures~\cite{ranta1975creep}, and the damage evolution in weathered limestone~\cite{torok2010current}. This body of work has yielded valuable constitutive models with experimentally calibrated parameters that could, in principle, enable accurate numerical simulations.

Yet in practice, this wealth of knowledge remains effectively inaccessible. The data exists as fragmented information scattered across thousands of PDF documents published in diverse journals spanning heritage science, civil engineering, materials science, and geotechnical engineering. Each publication presents its findings using varying nomenclature, different equation formats, and inconsistent parameter definitions. There is no unified database, no standardized schema, and no efficient means for a researcher to retrieve the specific constitutive model and parameters needed for their simulation task.

This ``data silo'' problem has tangible consequences. Researchers seeking material parameters must conduct extensive literature reviews---a process that is time-consuming, prone to oversight, and often incomplete. Alternatively, they resort to generic handbook values for ``brick'' or ``limestone'' that may bear little resemblance to the specific historical material under investigation. The result is either duplicated effort across research groups or simulations of questionable accuracy that could lead to misguided conservation decisions.

The obvious solution---automating the extraction of constitutive models from the scientific literature---faces formidable technical hurdles that have limited previous attempts, despite the growing momentum of materials informatics in other domains~\cite{tshitoyan2019unsupervised, kononova2019text, jablonka202314}.

First, the multimodal complexity of scientific documents poses significant challenges. Research papers are not merely text; they interleave prose, mathematical equations rendered in various formats (inline notation, displayed formulas, equation arrays), tables with numerical data, and figures with embedded information. Traditional Natural Language Processing (NLP) pipelines, designed primarily for plain text, struggle to parse and interpret this heterogeneous content~\cite{lopez2009grobid, blecher2023nougat, smock2022pubtables}.

Second, the identification of the governing constitutive law is complicated by the narrative structure of theoretical mechanics literature. Unlike structured data fields, a single research paper typically utilizes a multitude of equations to narrate a derivation process. These include precursor models introduced for comparison, intermediate derivation steps (e.g., thermodynamic consistency checks or yield surface evolution), and uncorrected trial formulations. Distinguishing the specific, calibrated constitutive model from this background of ``mathematical noise'' requires a high-level semantic understanding of the document's logical flow. An automated system must resolve long-range dependencies to determine which equation represents the author's final contribution rather than a literature review or a transient step in the derivation.

Third, and perhaps more fundamentally, is the challenge of symbolic grounding---the task of associating abstract mathematical symbols with their specific physical meanings. In constitutive modeling, the symbol $E$ might denote Young's modulus in one paper, activation energy in another, and total energy in a third. The Greek letter $\sigma$ typically represents stress, but its subscripts and the surrounding context determine whether it refers to normal stress, effective stress, yield stress, or something else entirely. Successfully extracting a constitutive model requires not just identifying the equation, but understanding what each symbol means~\cite{kristianto2014extracting}.

Fourth, contextual metadata is essential for proper application of extracted models. A stress-strain relationship for ``sandstone'' is of limited utility without knowing the specific type of sandstone, its provenance, the test conditions (temperature, humidity, strain rate, confining pressure), and the range of applicability. Capturing this contextual information requires understanding natural language descriptions that may be scattered across multiple sections of a paper.

Traditional approaches based on regular expressions, named entity recognition, or rule-based systems have achieved limited success in this domain. They can identify patterns but lack the reasoning capability to handle ambiguity, interpret context, and make inferences where information is implicit rather than explicit.

Recent advances in Large Language Models (LLMs) offer new possibilities for addressing these challenges. State-of-the-art models have demonstrated remarkable capabilities in zero-shot reasoning, understanding context, following complex instructions, and generating structured outputs~\cite{wang2023scientific, wei2022chain}. Crucially, these models can process and reason about mathematical notation, understand domain-specific terminology, and maintain coherence across long documents.

The few-shot and zero-shot learning capabilities of LLMs are particularly advantageous for domain-specific knowledge extraction tasks. Unlike traditional machine learning approaches that require large labeled training datasets---impractical to create for the specialized domain of heritage material constitutive models---LLMs can be guided through carefully designed prompts to perform extraction tasks with minimal or no task-specific examples~\cite{brown2020language, kojima2022large, dunn2022structured}.

However, applying LLMs to scientific knowledge extraction is not without challenges. LLMs are prone to ``hallucination''---generating plausible-sounding but factually incorrect information~\cite{ji2023survey}. Processing entire papers with large LLMs is computationally expensive, making brute-force approaches impractical for large-scale database construction. And ensuring consistent, structured outputs that conform to a predefined schema requires careful prompt engineering and validation mechanisms.

In this paper, we present an automated workflow for extracting mechanical constitutive models from scientific literature, specifically targeting cultural heritage materials. Our contributions are summarized as follows:

\begin{enumerate}
    \item \textbf{Development of a novel two-stage agentic framework.} We introduce a pipeline employing a cost-effective ``Gatekeeper'' agent for relevance screening and a high-capability ``Analyst'' agent for precise extraction~\cite{xi2023rise}. This collaborative design significantly reduces computational overhead while maintaining high precision in handling complex multimodal documents.
    
    \item \textbf{Establishment of a specialized mechanical constitutive model database.} Through the rigorous screening of over 2,000 papers, we identified 113 high-quality studies meeting strict criteria for heritage relevance and experimental validation. This effort yielded a structured repository containing 185 constitutive model instances and over 450 calibrated parameters.
    
    \item \textbf{Implementation of the Heritage Materials Constitutive Database Platform.} We developed a web-based system that facilitates intelligent data ingestion and semantic knowledge retrieval. This platform bridges the gap between static literature and engineering practice, enabling researchers to efficiently query range-bound parameters for downstream numerical simulations.
\end{enumerate}
\section{Methodology}\label{sec:method}

\subsection{Overview of the Agentic Framework}

We propose a novel Two-Stage Agentic Framework designed to automate the extraction of mechanical constitutive models from unstructured scientific literature. The framework addresses three primary challenges in this domain:(1) the high computational cost of processing large volumes of irrelevant documents~\cite{chen2023frugal}; (2) the complexity of isolating the target constitutive model amidst a high density of mathematical expressions, which often include intermediate derivation steps and uncorrected baseline models; and (3) the need for precise semantic grounding of mathematical symbols in diverse contexts.

\begin{figure}[htbp]
    \centering
    \includegraphics[width=0.95\textwidth]{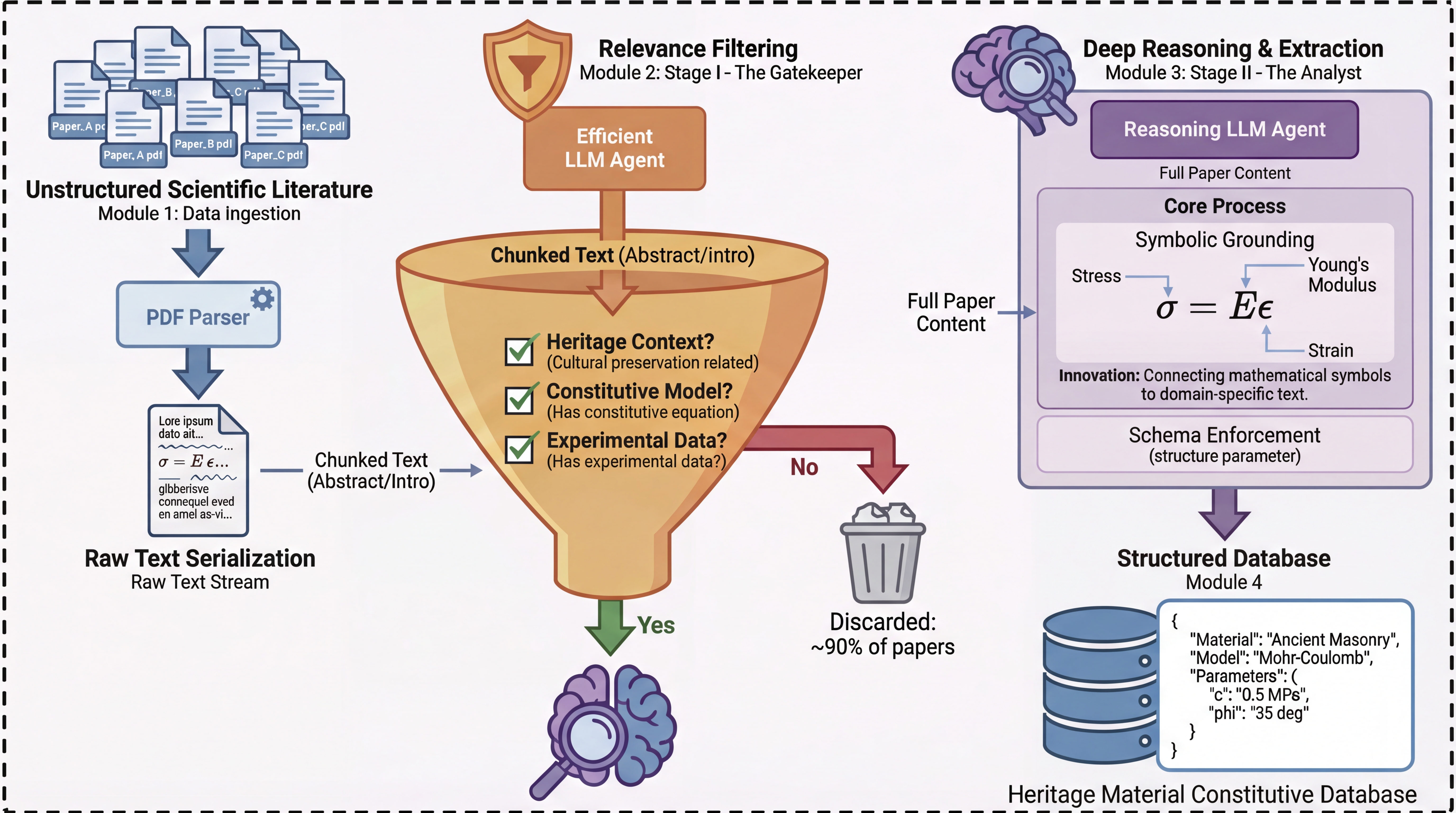}
    \caption{Overview of the Two-Stage Agentic Framework. The workflow adopts a coarse-to-fine strategy: (1) Raw PDF ingestion and serialization; (2) The Gatekeeper acts as a low-cost filter to discard irrelevant literature; (3) The Analyst employs deep reasoning for symbolic grounding; (4) The final parameters are stored in a structured JSON database.}
    \label{fig:framework}
\end{figure}

As illustrated in Figure~\ref{fig:framework}, the system architecture operates as a coherent pipeline comprising four functional modules. The workflow commences with the Data Ingestion Module, which parses raw PDF documents and serializes the heterogeneous content into a machine-readable text stream. This stream is then processed by Stage I (The Gatekeeper), a resource-efficient agent that acts as a relevance filter to discard non-pertinent literature based on broad-spectrum criteria. Surviving documents advance to Stage II (The Analyst), a high-reasoning capability agent that performs the core task of fine-grained extraction, utilizing symbolic grounding to map mathematical variables to their physical meanings. Finally, the validated outputs are consolidated into the Structured Database, where data is standardized according to a strict schema to ensure interoperability and ease of retrieval.

This hierarchical design follows the principle of ``coarse-to-fine'' processing, ensuring that computational resources are allocated efficiently by reserving high-capability models only for highly relevant documents~\cite{yue2023mammoth}.
% [Added Citation: Yue et al. (Mammoth/Coarse-to-fine) - Supports the hierarchical processing design]

\subsection{Data Acquisition and Preprocessing}

\subsubsection{Source Selection Criteria}
To construct a representative dataset for this study, we sourced research papers exclusively from the arXiv preprint repository, targeting categories relevant to physics and engineering. The retrieval process employed a systematic boolean search strategy designed to identify the intersection of heritage science and solid mechanics. We constructed search queries by combining keywords related to heritage materials with terms defining mechanical characterization and constitutive modeling. This initial retrieval phase focused on maximizing recall to ensure a comprehensive corpus for downstream filtering.

\subsubsection{PDF Parsing and Serialization}
Scientific papers in PDF format present a semi-structured data challenge. We utilized a custom parsing pipeline to serialize PDF documents into machine-readable text streams. To maintain the integrity of mathematical content, the parser was configured to retain raw character sequences, preserving inline notations where possible.

For the subsequent Stage I processing, we implemented a Head-Truncation Strategy. Since key metadata regarding a paper's scope and methodology are typically concentrated in the opening sections, we extracted only the initial segment (approximately 8,000 characters) of each document. This strategy ensures that the Gatekeeper agent receives sufficient context for relevance judgment without incurring the computational cost of processing full-text body paragraphs.

\subsection{Stage I: The Gatekeeper Agent (Relevance Filtering)}

The first stage of our framework employs a resource-efficient Large Language Model designated as the ``Gatekeeper.'' Its primary objective is to act as a semantic filter, discarding documents that do not strictly align with the database's inclusion criteria. This hierarchical filtering prevents the downstream ``Analyst'' agent from processing irrelevant literature.

The Gatekeeper operates under a constrained reasoning protocol, evaluating each document against three non-negotiable dimensions. First, the agent verifies Domain Relevance to ensure the study focuses explicitly on cultural heritage materials rather than modern synthetic composites. Second, it checks for Theoretical Content, confirming that the paper formulates or utilizes a specific mechanical constitutive model. Third, the agent mandates Experimental Validation, requiring the presence of empirical data used for parameter calibration. Only documents satisfying all three conditions are classified as relevant and forwarded to the full-text processing pipeline.

\subsection{Stage II: The Analyst Agent (Knowledge Extraction)}

Papers validated by the Gatekeeper are processed by the ``Analyst'' agent, a module powered by a foundation model with advanced reasoning capabilities. Unlike the binary classification task in Stage I, this stage involves a complex sequence-to-structure generation task. We formalize this process as mapping an unstructured document $\mathcal{D}$ to a structured knowledge graph $\mathcal{G}$ under a set of semantic constraints.

\subsubsection{Context-Aware Target Identification}
A critical challenge in mechanical literature is the ``dense mathematical landscape.'' A single theoretical paper typically contains a multitude of equations, ranging from fundamental physical laws (e.g., conservation of mass) to intermediate derivation steps and unmodified legacy models used for comparative analysis. 
Distinguishing the authors' proposed constitutive model from these auxiliary equations is non-trivial and cannot be achieved through visual formatting features alone. It requires a high level of contextual understanding to interpret the narrative flow—specifically, identifying linguistic cues that signal the transition from derivation to the final proposal (e.g., ``we modify the standard Maxwell model by introducing...''). The Analyst agent is explicitly designed to navigate this complexity, filtering out derivational intermediates and baseline formulas to pinpoint the definitive constitutive relation.

\subsubsection{Mathematical Formalization of Symbolic Grounding}
The core challenge in extracting constitutive models is Symbolic Grounding---the disambiguation of mathematical notation based on context~\cite{schubotz2016semantification}. A constitutive equation $\mathcal{E}$ extracted from the text typically contains a set of abstract symbols $S = \{s_1, s_2, \dots, s_n\}$. In isolation, a symbol $s_i$ is semantically ambiguous.
% [Added Citation: Schubotz et al. - Specific paper on semantic challenges in mathematical notation]

For example, the symbol $\phi$ could denote "porosity" in a petrophysical study or "internal friction angle" in a geotechnical context. We define the grounding task as finding a bijective mapping function $f: S \to P$, where $P$ is the set of physical definitions derived from the local context $C_{local}$. The Analyst agent is optimized to maximize the conditional probability of the correct mapping:

\begin{equation}
    \hat{P} = \operatorname*{argmax}_{P \in \mathcal{P}} \ p(P \mid S, \mathcal{E}, C_{local})
\end{equation}

By strictly enforcing this probabilistic mapping, the system ensures that generic variables are accurately bound to their specific rheological meanings (e.g., mapping $\sigma_c$ specifically to "Uniaxial Compressive Strength" rather than generic stress).

\subsubsection{Schema-Constrained Decoding}
To transform the extracted knowledge into a queryable format, we enforce a Schema-Constraint Decoding strategy~\cite{willard2023efficient}. Instead of free-form generation, the model's output space is restricted to a pre-defined JSON schema $\mathcal{J}$. This schema mandates the extraction of a 5-tuple vector for each material model:
% [Added Citation: Willard et al. - The technical paper for constrained generation/JSON mode]

\begin{equation}
    V_{\text{model}} = \langle \mathcal{E}, \mathcal{V}_{\text{map}}, \mathcal{M}_{\text{meta}}, \Pi_{\text{params}}, \mathcal{S}_{\text{val}} \rangle
\end{equation}
where $\mathcal{E}$ is the constitutive equation in standard LaTeX format, $\mathcal{V}_{\text{map}}$ represents the symbol-definition pairs, $\mathcal{M}_{\text{meta}}$ denotes material metadata (e.g., "Ancient Sandstone"), $\Pi_{\text{params}}$ contains numerical parameter values, and $\mathcal{S}_{\text{val}}$ identifies the experimental validation method.

\subsubsection{Error Handling and Self-Correction}
To mitigate the inherent non-determinism of generative models, we implemented a closed-loop verification mechanism. The system parses the output string against the strict JSON schema $\mathcal{J}$. Let $O_t$ be the output at step $t$. If a syntax error or schema violation $\epsilon$ is detected (i.e., $O_t \notin \mathcal{J}$), the error signal $\epsilon$ is fed back into the model context as a negative constraint. The model then performs a self-correction step to generate $O_{t+1}$, significantly improving the rigorousness of the final database.
\section{Results}\label{sec:results}

\subsection{Database Construction Statistics}

To rigorously test the scalability of our framework, we processed an initial corpus of over 2,000 research papers sourced from the arXiv preprint repository. The Stage I (Gatekeeper) agent acted as a high-throughput filter, rejecting the vast majority of documents identified as irrelevant.

This strict filtering yielded a refined set of 113 highly relevant papers. From these core documents, the Stage II (Analyst) agent successfully populated the Heritage Material Constitutive Database. Specifically, the system extracted 185 validated constitutive model instances and over 450 experimentally calibrated material parameters. Crucially, each entry is linked with rich contextual metadata, including material provenance and environmental testing conditions, ensuring the data is ready for high-fidelity numerical simulation.

\begin{figure}[htbp]
    \centering
    \includegraphics[width=0.6\textwidth]{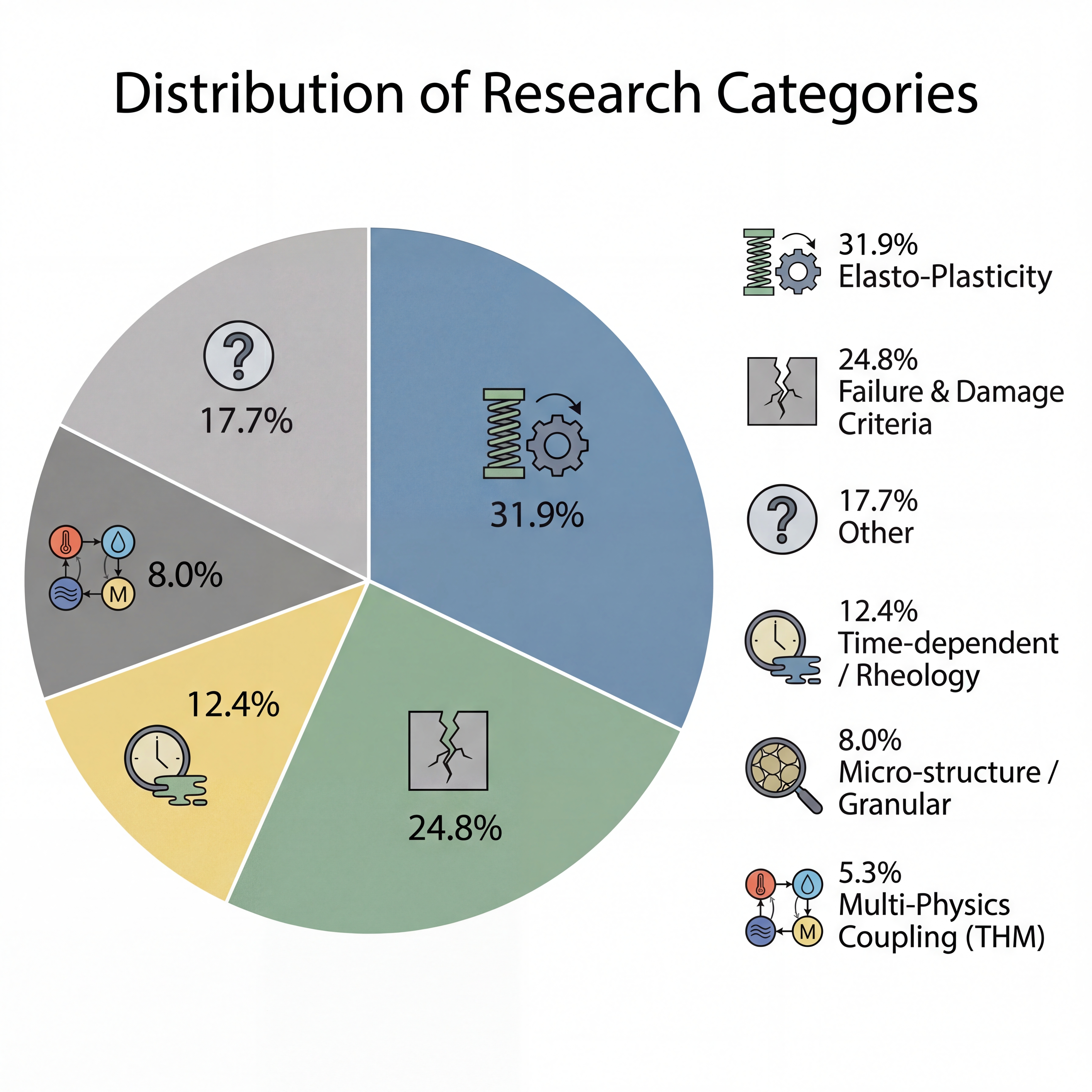} 
    \caption{Distribution of Constitutive Mechanisms. The framework successfully categorized extracted models into distinct rheological behaviors. The prevalence of plasticity and damage models aligns with the need for safety assessment in heritage structures.}
    \label{fig:statistics}
\end{figure}

Figure~\ref{fig:statistics} illustrates the distribution of the extracted constitutive models by mechanical mechanism. The dataset is dominated by Elasto-Plasticity models (31.9\%) and Failure \& Damage Criteria (24.8\%), reflecting the heritage conservation community's primary focus on assessing structural stability and collapse risks. Other significant categories include Time-dependent/Rheology models (12.4\%), which are critical for analyzing long-term creep in historical timber and masonry.

\subsection{Performance Evaluation}

To quantitatively assess the extraction quality, we established a rigorous Ground Truth (GT) dataset. The entire set of 113 relevant papers identified in Stage I was manually annotated by domain experts in solid mechanics. This manual annotation process identified a total of 222 target constitutive model entities (Total Ground Truth), which served as the gold standard for validation.

We defined the evaluation metrics as follows: A True Positive (TP) occurs when the agent correctly extracts a constitutive model that matches the expert annotation in both equation form and physical meaning. A False Positive (FP) represents an extracted entry that is either hallucinated or incorrectly identified (e.g., extracting an intermediate derivation step instead of the final model). A False Negative (FN) indicates a valid model present in the text that the agent failed to capture.

Based on these definitions, the framework demonstrated robust capabilities with a Precision of 80.4\% ($TP/[TP+FP]$), a Recall of 83.3\% ($TP/[TP+FN]$), and an overall F1-Score of 81.9\%. 

\begin{figure}[htbp]
    \centering
    \includegraphics[width=0.9\textwidth]{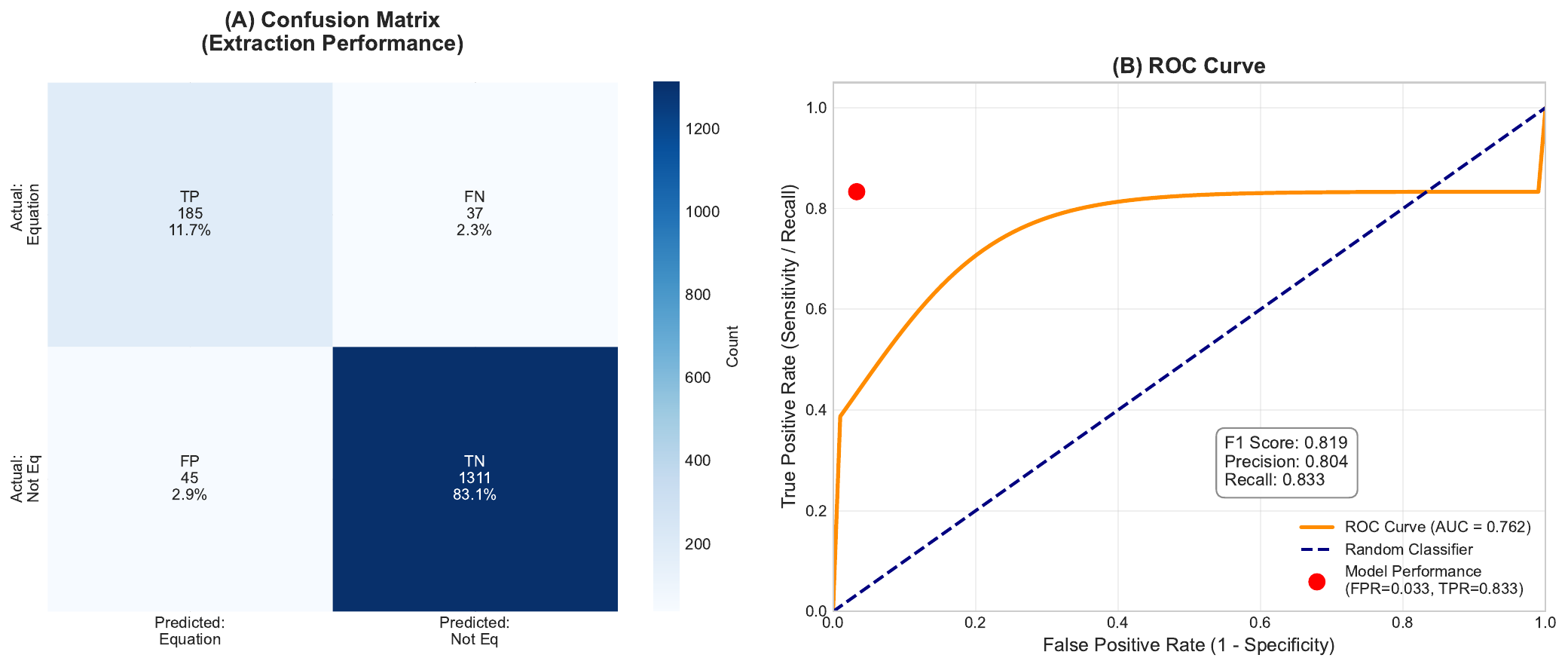} 
    \caption{Quantitative Evaluation of the Framework. (a) Confusion Matrix showing the extraction performance (TP=185, TN=1311); (b) ROC Curve with an AUC of 0.782. The selected operating point (Red Dot) corresponds to a low False Positive Rate (3.3\%), prioritizing the reliability of database entries.}
    \label{fig:evaluation}
\end{figure}

In addition to these aggregate metrics, Figure~\ref{fig:evaluation} visualizes the detailed performance characteristics through a binary classification analysis.

The Confusion Matrix (Fig.~\ref{fig:evaluation}a) reveals a True Positive (TP) count of 185 against a background of 1311 True Negatives (TN). Note that in this context, TNs represent potential candidate text blocks or equations that were correctly rejected by the agent as non-constitutive models. The model maintains a balanced error distribution between False Positives (45) and False Negatives (37). A retrospective analysis suggests that these residual errors are largely attributable to the parsing challenges of non-standard tabular structures in older PDFs, rather than semantic reasoning failures.

The ROC curve (Fig.~\ref{fig:evaluation}b) further characterizes the classifier's discriminatory power, yielding an Area Under Curve (AUC) of 0.782. Notably, the operational point of our deployed model (indicated by the red dot) is situated at a False Positive Rate (FPR) of only 3.3\%. This low FPR is strategically vital for automated database construction, as it minimizes the contamination of the repository with irrelevant noise, prioritizing data purity even at the slight expense of recall.

The results indicate that while the system is highly effective (F1 $\approx$ 0.82), a small fraction of extracted data benefits from human verification. Nevertheless, compared to fully manual literature review, this semi-automated workflow reduces the workload by orders of magnitude while maintaining a high standard of data integrity.

\subsection{Qualitative Analysis: Reasoning Capabilities}

\begin{figure}[htbp]
    \centering
    \includegraphics[width=0.95\textwidth]{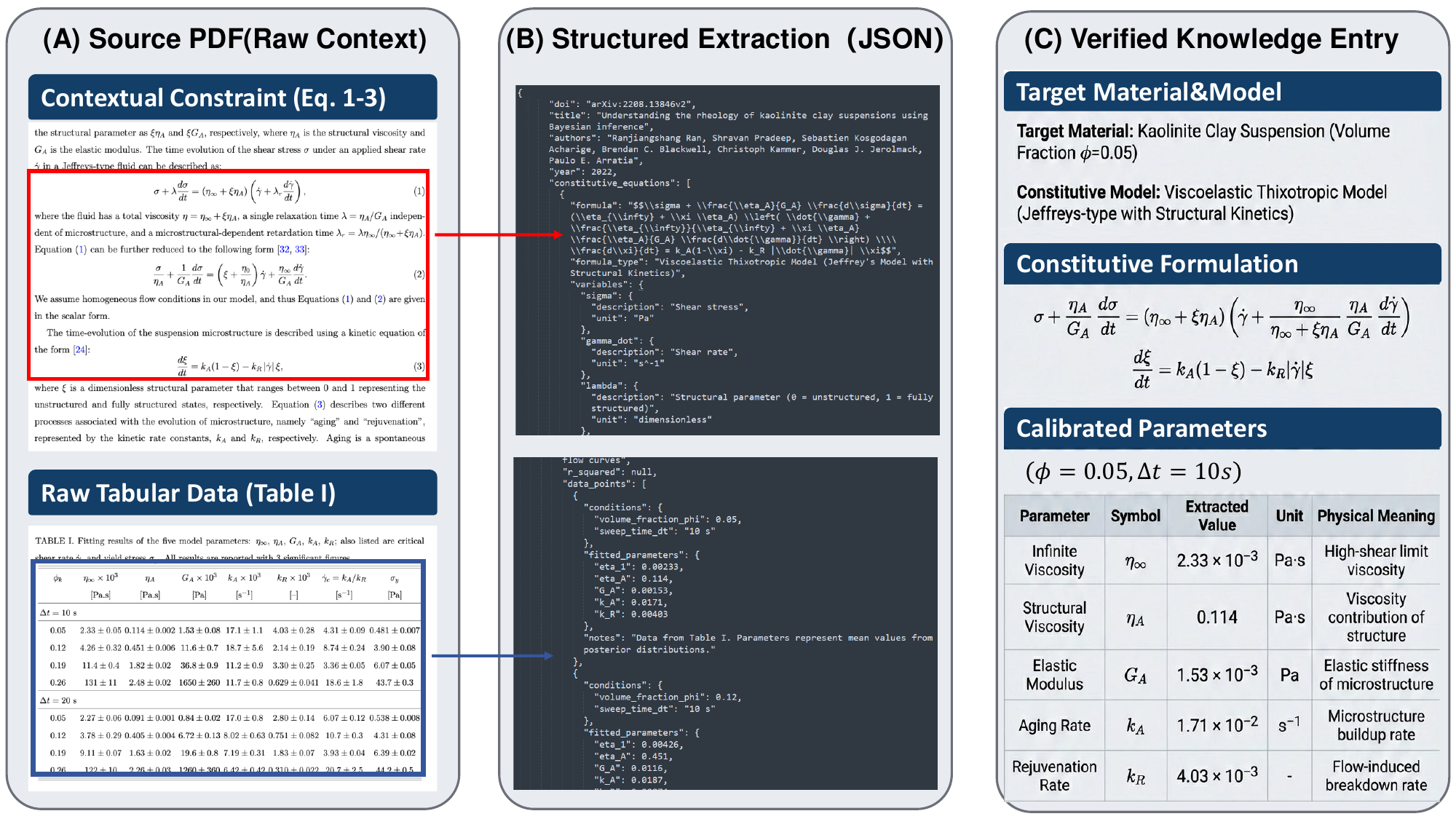}
    \caption{Qualitative Extraction Case Study. The extraction process for a Jeffreys-type Viscoelastic model. (A) The raw PDF presents the constitutive law components scattered across disjointed sections and data in a complex table. (B) The Agent fuses these components into a structured JSON while filtering derivation noise. (C) The final Verified Knowledge Entry shows correct Symbolic Grounding (mapping $\xi$ to structural kinetics) and Physical Plausibility Verification (resolving the ambiguous header scale via domain constraints).}
    \label{fig:extraction_case}
\end{figure}

Beyond statistical metrics, the qualitative capability of the framework is visually demonstrated in the case study of Figure~\ref{fig:extraction_case}. This example, extracted from a study on Kaolinite clay rheology~\cite{ran2023understanding}, highlights two critical reasoning advantages of the Agentic framework over traditional rule-based parsers.

First, the system demonstrates sophisticated Contextual Fusion and Noise Filtering. In scientific literature, constitutive models are rarely defined in a single contiguous block; instead, they are often distributed across disjointed sections, intermixed with derivation steps.
As seen in Figure~\ref{fig:extraction_case} (Panel A), the source document spatially separates the stress evolution component from the structural kinetic law.
Crucially, the text is also interspersed with intermediate derivation forms (e.g., uncorrected precursor definitions).
Traditional extractors would blindly capture all mathematical blocks or treat them as unrelated strings.
In contrast, our Analyst agent performed Governing Equation Discrimination: it successfully filtered out the intermediate mathematical noise and aggregated the spatially distinct components into a single, self-contained constitutive formulation shown in the Verified Knowledge Entry (Panel C).

Second, the agent demonstrates remarkable Domain-Adaptive Disambiguation and Physical Plausibility Reasoning.
A persistent challenge in scientific extraction is the syntactic ambiguity of tabular headers.
In Table I (Panel A), the column header ``$\eta_\infty \times 10^3$'' is notationally ambiguous: it could imply a magnitude of $10^3$ (common in geotechnical soil mechanics) or a scaling factor requiring multiplication by $10^{-3}$ (to display small values legibly).
A rule-based parser typically fails to resolve this without hard-coded heuristics.
However, our Analyst agent successfully resolved this by grounding the extraction in the specific physical domain of the paper---``aqueous clay suspensions'' rather than consolidated soils.
Recognizing that suspension viscosities physically adhere to the order of water ($\sim 10^{-3}$ Pa$\cdot$s), the agent correctly interpreted the header as a scaling factor, converting the raw value $2.33$ to the standardized $2.33 \times 10^{-3}$ Pa$\cdot$s.
This proves the framework's ability to utilize semantic context to impose physical constraints on extracted data, avoiding order-of-magnitude hallucinations.
\section{Application Demonstration}\label{sec:application}

To bridge the gap between unstructured literature and engineering practice, we developed the Heritage Materials Constitutive Database Platform. This web-based system supports a closed-loop workflow, enabling both the intelligent ingestion of new data and the semantic retrieval of existing knowledge.

\subsection{Intelligent Data Ingestion}

\begin{figure}[htbp]
    \centering
    \includegraphics[width=0.9\textwidth]{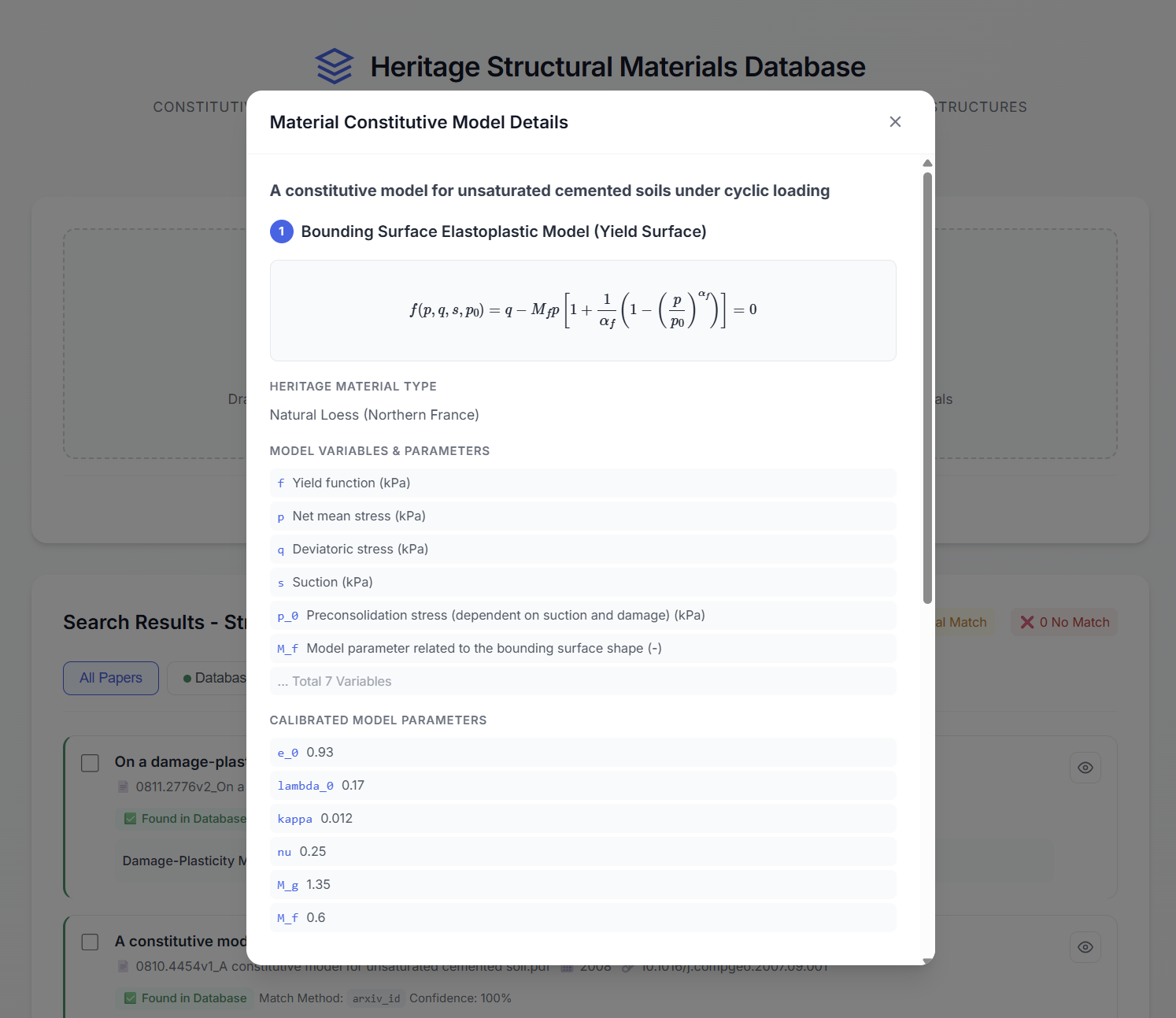}
    \caption{Automated Data Ingestion Interface. The detail view shows the extraction results from an uploaded PDF. The system automatically identifies the constitutive model, renders the equation, and populates the fitted parameters for user verification.}
    \label{fig:ingestion}
\end{figure}

The platform features an automated ingestion module designed to streamline the contribution of new data. As demonstrated in Figure~\ref{fig:ingestion}, researchers can simply upload raw PDF literature via the web interface. Upon upload, the backend extraction agents immediately initiate a parsing sequence. First, the system identifies the specific constitutive model and renders its mathematical formulation using LaTeX for visual verification. Simultaneously, it extracts the fitted parameters—such as cohesion and friction angle—and automatically populates the corresponding database fields. This workflow effectively transforms static, unstructured PDFs into structured digital assets without requiring manual data entry.

\subsection{Semantic Knowledge Retrieval}

\begin{figure}[htbp]
    \centering
    \includegraphics[width=0.9\textwidth]{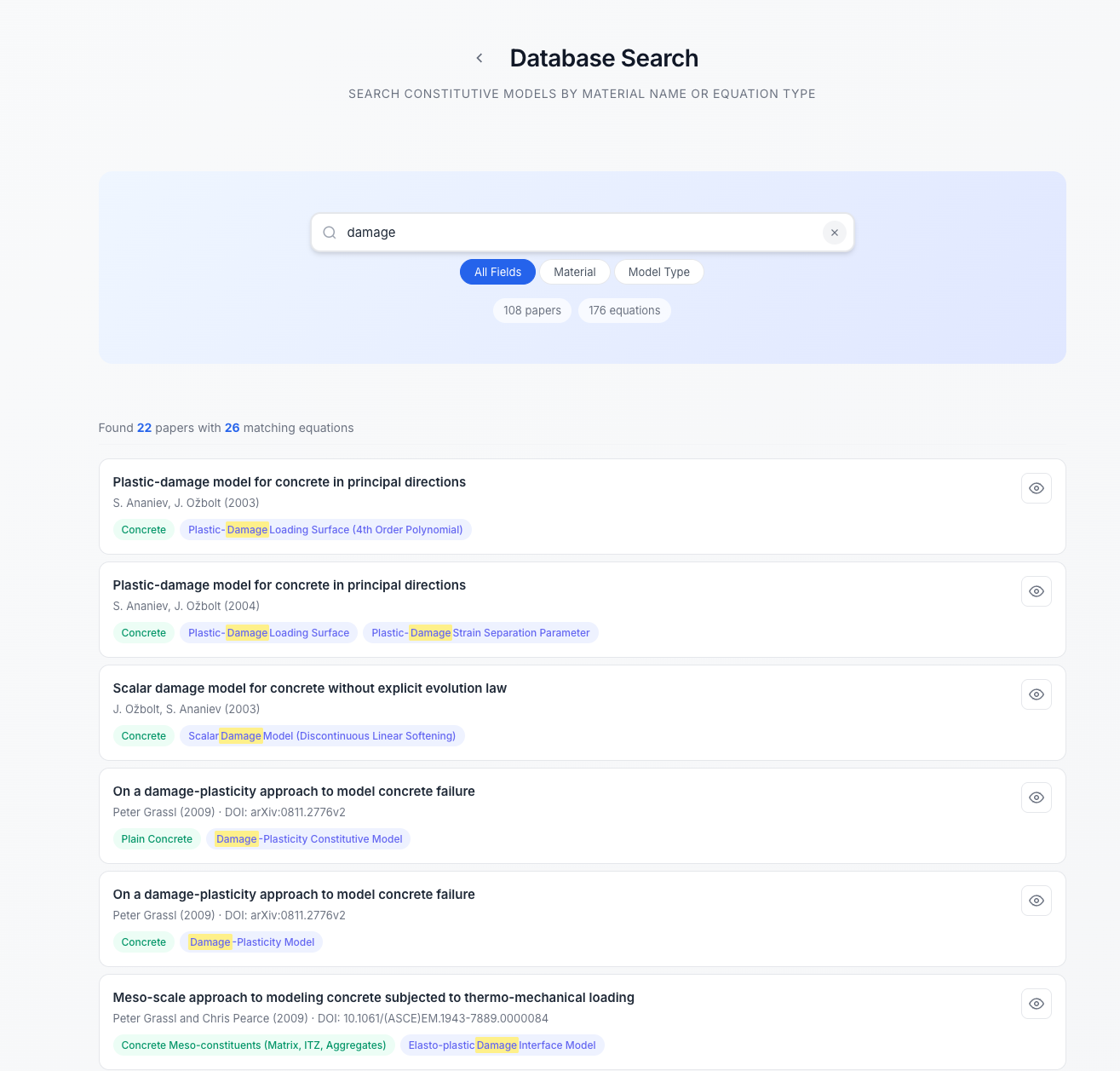}
    \caption{Semantic Retrieval Interface. The search module allows researchers to filter constitutive models by material type or property. The results list displays key metadata, enabling rapid selection of appropriate parameters for simulation.}
    \label{fig:retrieval}
\end{figure}

Beyond individual file processing, the platform facilitates data discovery through a dedicated search engine. As illustrated in Figure~\ref{fig:retrieval}, users can query the aggregated database by specific material categories or mechanical properties. In response to a query, the system retrieves all matching entries, displaying the calibrated parameters alongside their source validation status. Consequently, this capability allows conservation engineers to rapidly establish range-bound reference values for numerical simulations, significantly reducing the lead time compared to conducting exhaustive manual literature reviews.

\section{Discussion}\label{sec:discussion}

The results of this study mark a departure from the traditional literature review workflow in heritage conservation. While manual compilation is accurate but labor-intensive, our Two-Stage Agentic Framework offers a scalable alternative. It is important to acknowledge that the system achieves an F1-score of approximately 0.82, not perfection. However, in an engineering context, this creates a valuable "Human-in-the-loop" workflow: the Agent acts as a high-speed pre-screener that reduces approximately 90\% of the manual workload, leaving domain experts to verify the "final mile" of the data (as demonstrated in the verified entries in Section 4). This synergy combines the scalability of AI with the reliability of human expertise, making large-scale database construction feasible for the first time.

Furthermore, our approach specifically addresses the unique "Data Poverty" challenge in heritage science. Unlike general computer vision domains (e.g., ImageNet) that benefit from millions of data points, valuable heritage experimental data is scarce and scattered across decades of PDF publications. By structuring 185 constitutive models from 113 core papers, we have aggregated what is likely the largest structured repository of heritage mechanical properties to date. This transition from scattered PDFs to a unified SQL database enables researchers to move from heuristic guessing to data-driven probabilistic assessment, such as determining the confidence intervals of material stiffness.

Despite these advancements, challenges remain in handling extreme document variability. First, regarding document quality, the parser exhibits sensitivity to the layout of older, low-resolution scanned documents (pre-2000s), where significant noise can disrupt structural recognition. Second, regarding multimodal extraction, while our current agent utilizes Vision-Language capabilities to identify and interpret figures, the precise digitization of continuous data points from complex scientific charts (e.g., dense stress-strain hysteresis loops) remains a non-trivial task compared to textual parameter extraction. Future work will focus on enhancing the agent's ability to perform "Chart-to-Data" conversion, creating a truly holistic extraction pipeline that fully digitizes both semantic and graphical information.

\section{Conclusion}\label{sec:conclusion}

This paper presented an automated, agentic workflow for extracting mechanical constitutive models from scientific literature, explicitly tailored to address the "Data Silo" problem in cultural heritage conservation.

By leveraging a cost-effective Gatekeeper-Analyst architecture, we successfully filtered a corpus of over 2,000 papers to isolate 113 high-relevance documents. The system extracted 185 constitutive model instances and over 450 calibrated parameters with a precision of 80.4\%. Crucially, the framework overcomes the challenge of symbol ambiguity through context-aware symbolic grounding, ensuring that mathematical variables are correctly mapped to their physical definitions.

The utility of this framework is realized through the developed Heritage Materials Constitutive Database Platform, which transforms static literature into an interactive, searchable digital asset. As demonstrated in our application scenarios, this platform enables rapid retrieval of material parameters for computational modeling and structural assessment. We believe this work lays the foundation for the "Digital Material Twin" of our built heritage, ensuring that the scientific knowledge of the past is preserved and operationalized for future conservation efforts.

\section*{Acknowledgments}
This work is supported by the Advanced Materials-National Science and Technology Major Project(2025ZD0618802) and the Shanghai Technical Service Center of Science and Engineering Computing, Shanghai University.

\backmatter

\bibliography{sn-bibliography}

\end{document}